\documentclass[twocolumn,secnumarabic,amssymb, aps, prl]{revtex4-1}
\usepackage{graphicx}

\usepackage{color}

\begin{document}

\title{Structural Origin of the Two-Step Glass Transition}%
\author{Xinzhuo Liu$^{1,2}$}
\author{Huaguang Wang$^{1,2}$}
\email{hgwang@suda.edu.cn}
\author{Zexin Zhang$^{1,2}$}
\email{zhangzx@suda.edu.cn}
\author{Xinsheng Sean Ling$^{3}$}
\email{xinsheng$\_$ling@brown.edu}
\affiliation{$^{1}$Institute for Advanced Study, Center for Soft Condensed Matter Physics and Interdisciplinary Research, School of Physical Science and Technology, Soochow University, Suzhou 215006, China}
\affiliation{$^{2}$College of Chemistry, Chemical Engineering and Materials Science, Soochow University, Suzhou 215123, China}
\affiliation{$^{3}$Department of Physics, Brown University, Providence, RI 02912, USA}
\date{\today}%

\begin{abstract}
The glass transition is a long-standing problem in physics. Identifying the structural origin of the transition may lead to the ultimate solution to the problem. Here, for the first time, we discover such a structural origin by proposing a novel method to analyze structure-dynamics relation in glasses. An interesting two-step glass transition, with rotational glass transition preceding translational one, is identified experimentally in colloidal rod systems. During the transition, parallel and perpendicularly packed rods are found to form local free energy minima in configurational space, separated by an activation barrier. This barrier increases significantly when rotational glass transition is approached; thereby the rotational motion is frozen while the translational one remains diffusive. We argue that the activation barrier for rotation is the origin of the two-step glass transition. Such an activation barrier between well-defined local configurations holds the key to understand the two-step glass transition in general.
\end{abstract}

\maketitle

Identifying the origins of the glass transition remains one of the most challenging problems in condensed matter physics and materials science \cite{Angell1995,Debenedetti2001,Berthier2011,Weeks2000}. Inspired by the successful identification of topological defects, edge dislocations and disclinations, in the equilibrium phase transitions of 2D crystallization and melting \cite{Kosterlitz1973,Halperin1978,Young1979}, researchers have long searched for the structural mechanisms in the glass transition with limited success \cite{Royall2015,Tanaka2019}. Numerous experiments and simulations of glass-forming liquids suggest the importance of locally ordered structures \cite{Spaepen2000,Shintani2006,Kawasaki2007,Hu2015,Tanaka2019}. They have been suggested to be responsible for the dynamics arrest associated with a glass transition. Since the deciding structural signature in glass transition remains elusive, the roles of local order structures on the dynamics remain controversial and are still actively studied \cite{Tanaka2019,Tong2018}.
Recently there are growing interests in the glass transition in systems of anisotropic colloids  \cite{chong2002pre, Pfleiderer2008, kob2009prl, Kramb2010,Zheng2011,Yunker2011,Mishra2013,kang2013prl}.  For example it has been suggested that local nematic structures may provide a mechanism for the glass transition of colloidal ellipsoids  \cite{Zheng2014,Mishra2013,Yunker2011}.  Specifically, for ellipsoids with large aspect ratio $p > 2.5$, it has been proposed that the formation of pseudo-nematic domains leads to a decoupling between translational and orientational motion \cite{Zheng2011}. Consequently, the ellipsoid system undergoes an interesting two-step glass transition.  It forms a glass in orientational degrees of freedom at a lower density followed by a glass transition in translational degrees of freedom at a higher density  \cite{Zheng2011}. In the orientational glass, cooperative rearrangements in orientational motion are prohibited, while the particles still display collective translational motion \cite{Zheng2011}.  For $p < 2.5$, the pseudo-nematic domains disappear,  the two glass transitions merge into a one-step transition \cite{Mishra2013,Zheng2014,Schilling1997}. Such local order is also observed in ellipsoids with attractive interaction even for $p < 2.5$, where the pseudo-nematic domains are attributed to the attraction \cite{Mishra2013}. It appears that the two-step glass transition may be related to the pseudo-nematic domains. However, a quantitative description of the local order structure is still missing, and a universal structural mechanism for the two-step glass transitions remains elusive.

Here we report an experimental study of the glass transition in 2D suspensions of colloidal rods \cite{Liu}. Like ellipsoids, colloidal rods undergo a two-step glass transition even for small aspect ratio, $p$ = 1.5. Unlike ellipsoids, there is no evidence of pseudo-nematic domain formation at any packing density. We extracted the free energy landscape of the rod-rod interactions from statistical analyses of the relative angle distribution between the rods. We found that the parallel and perpendicular local orientational configurations of the rods are local free energy minima, separated by an activation barrier. The large activation barriers between the local configurations hinder rotation of the rods, resulting in a decoupling between translational and orientational motion, and consequently a two-step glass transition. We believe that the activation barrier between well-defined local configurations holds the key for a universal understanding of the two-step glass transition.

\begin{figure*} [tbhp]
	\includegraphics[width=1.95\columnwidth]{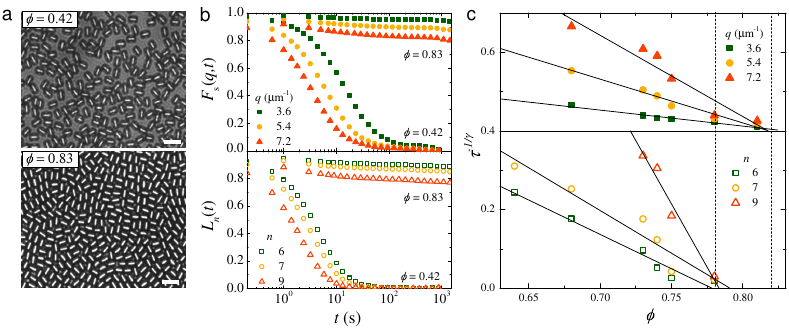}
	\caption{\textbf{Relaxation times show a two-step glass transition.} (a) Brightfield micrographs of the colloidal rod monolayers. Only part of the field of the view is shown for clarity. Scale bar: 10 $\mu m$. (b) $F_s(q,t)$ for different $q$ (top panel) and $L_n(t)$ for different $n$ (bottom panel) at $\phi = 0.42$ and $\phi = 0.83$. (c) The fitted relaxation time $\tau(\phi)\sim(\phi_g-\phi)^{-\gamma}$. $\gamma$ = 1.77 and 1.83 for translational (upper panel) and orientational (lower panel) relaxation times, respectively. The $\tau^{-1/\gamma}$ for translational relaxation time is shifted by 0.4 for clarity. The scalings show there are two glass transitions: $\phi_g^\theta = 0.78\pm0.01$ for rotational motion and $\phi_g^T = 0.82\pm0.01$ for translational motion (marked by the two vertical dotted lines).}
\end{figure*}

Our colloidal rods were fabricated by stretching non-cross-linked polystyrene (PS) spheres \cite{Champion2007}. We prepared three batches of rods with aspect ratio $p = L/D =  5.1\,\mu m/3.4\,\mu m = 1.5$, $p =  6.1\,\mu m/2.7\,\mu m = 2.3$ and $p =  6.8\,\mu m/2.5\,\mu m = 2.7$, where $L$ and $D$ are the length and diameter of the rods, respectively. The polydispersity of the aspect ratio was 5-10\% for all $p$'s. The suspension of the rods was confined between two glass coverslips to produce a monolayer of the particles (Fig. 1a). The areal fraction $\phi = A\rho$, where \emph{A} is the cross sectional area of the rod and $\rho$ is the particle number density. The thermal motion of the rods was recorded by an optical microscope with a CCD camera at 5 frames per second for low $\phi$ and 1 frame per second for high $\phi$. During measurements at each $\phi$, no drift was observed. The center positions and orientations of individual rods were obtained using the ImageJ. The angular resolution was $1^\circ$, and the spatial resolution was 60 nm. More experimental details are in the Supplemental Information.

Our experimental system consists of colloidal rods of different anisotropies and varying packing density confined between two glass coverslips. The system is a 2D liquid at low densities and a glass at high densities. In order to accurately characterize the glass transition, we applied two different approaches, the mode-coupling theory (MCT) \cite{Das2004}, and the Vogel-Tammann-Fulcher (VTF) model \cite{Keys2007}. In this work, we have chosen the MCT scaling over the VFT scaling in order to compare the results with those for ellipsoidal glass \cite{Zheng2011,Mishra2013}, where the MCT scaling is found to give better description of heterogeneous glassy dynamics. A typical result using VTF is included in the supplementary information (Fig. S2). The MCT approach of characterizing a glass transition in colloidal systems is to extract translational and rotational relaxation times as a function of packing density.  Here, the self-intermediate scattering function $F_s(q,t) \equiv \frac{1}{N}\langle \sum^{N}_{j=1}e^{i\mathbf{q}\cdot[\mathbf{r}_j (t)-\mathbf{r}_j (0)]} \rangle$ and orientation correlation function $L_n(t) \equiv \frac{1}{N}\langle\sum^{N}_{j=1}$cos$n[\theta_j(t)-\theta_j(0)] \rangle$ were calculated, where $\mathbf{r}_j (t)$ and $\theta_j(t)$ are the position and orientation of rod \emph{j} at time \emph{t}, \emph{N} is the total number of particles, $\mathbf{q}$ is the scattering vector, \emph{n} is a positive integer, and $<>$ denotes a time average. The relaxation time $\tau$ is defined as the time at which $F_s(q,t)$ and $L_n(t)$ have decayed to 1/e \cite{Kob1994,Zheng2011,Mishra2013}. Note that $F_s(q,t)$ and $L_n(t)$ decay faster for larger value of \emph{q} and \emph{n}, respectively (Fig. 1b), and different choices of \emph{q} and \emph{n} can yield the same glass transition point \cite{Zheng2011,Mishra2013}.

The glass transition point $\phi_g$ is determined from the MCT scalling, following previous work on the glass transition of colloidal ellipsoids \cite{Zheng2011,Zheng2014,Mishra2013}. According to the MCT, the relaxation time $\tau$ diverges algebraically at $\phi_g$: $\tau(\phi)\sim(\phi_g-\phi)^{-\gamma}$, where $\gamma= \frac{1}{2a} + \frac{1}{2b}$ \cite{Gotze1992}. Here $a$ and $b$ are exponents in the critical-decay law $F_s(q,t) = f_q + h_qt^{-a}$ and the von Schweidler law $F_s(q,t) = f_q-h_qt^{-b}$ for the initial $\beta$ relaxation and the crossover time to the $\alpha$ relaxation, respectively \cite{Bayer2007,Hajnal2009,Pfleiderer2008}. $f_q$ and $h_q$ are the plateau height and amplitude. Here we obtain $b$ from fits to $F_s(q, t)$ and $L_n(t)$, and $a$ from Ref. \cite{Gotze1992}. The result shows that the $\tau^{-1/\gamma}$ is linear in $\phi$ for different choices of \emph{q} and \emph{n}, and expectedly, all the scalings show there are two glass transitions at $\phi_g^\theta = 0.78\pm0.01$ for rotational motion and $\phi_g^T = 0.82\pm0.01$ for translational motion for rods with $p = 1.5$.  Thus the orientational glass transition occurs at a lower density than the translational glass transition (Fig. 1c). We also have confirmed such two-step glass transition by determining the glass transition points from the Vogel-Tammann-Fulcher (VTF) scaling \cite{Keys2007} (see SI, Fig. S2). Moreover, this two-step glass transition is also observed for rods with $p$ = 2.3 and 2.7 (see SI, Fig. S5).

\begin{figure} 
	\includegraphics[width=0.9\columnwidth]{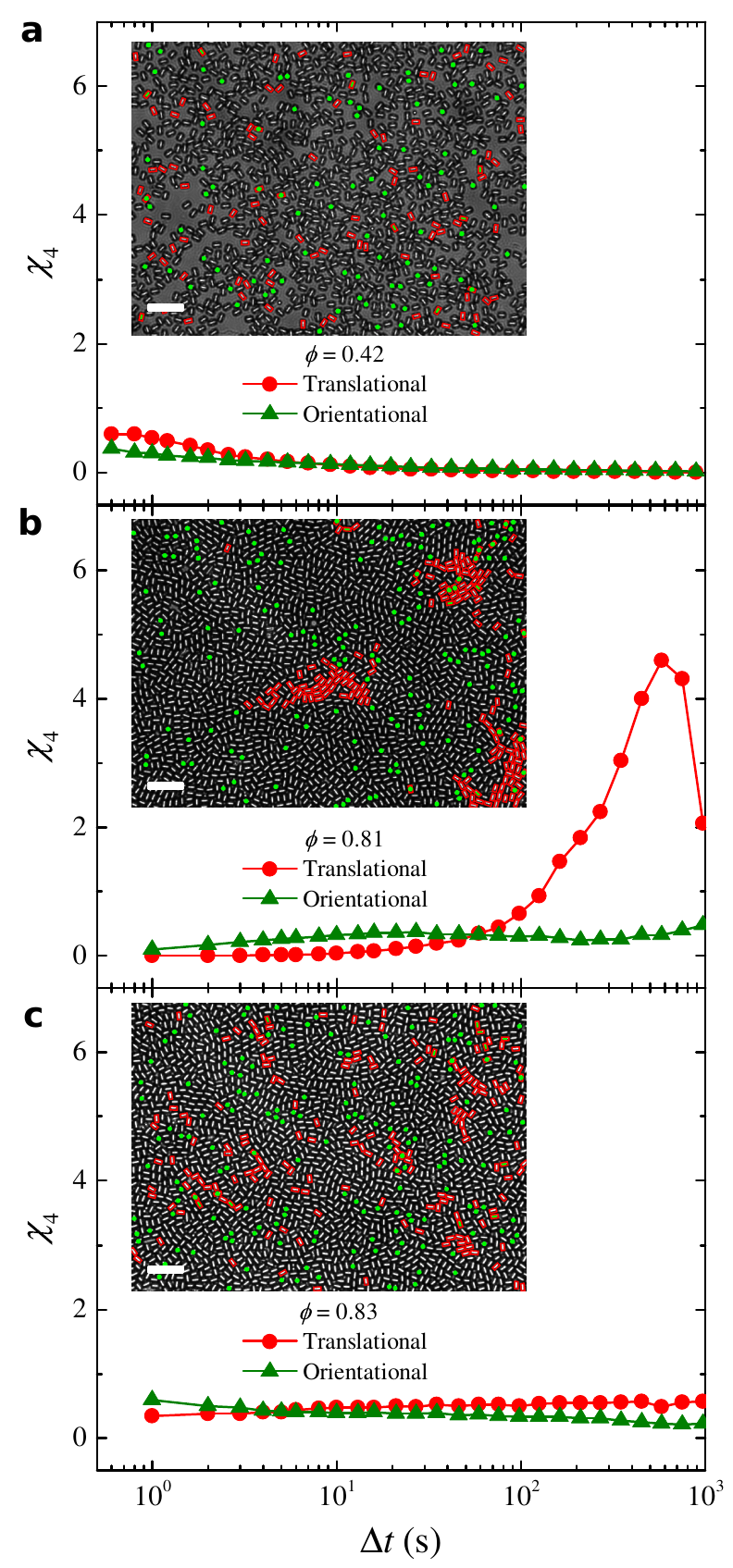}
	\caption{\label{fig2} \textbf{Dynamical heterogeneity confirms the existence of the orientational glass state.} The four-point susceptibility, $\chi_4$, for translational motion and orientational motion for the rods with aspect ratio, $p = 1.5$. Insets are bright-field micrographs of fast-moving particles for translational motion (red squares) and orientational motion (green dots) at, from top to bottom, $\phi = 0.42, 0.81 and 0.83$.The lack of orientational cooperative motion and the appearance of strong translational cooperative motion at $\phi$ = 0.81 indicate that the orientational motion is frozen but the translational motion remains., confirming the existence of the orientational glass phase. The fast-moving particles are defined as the particles with the 10\% largest displacements over the lag time $\Delta t$ that maximizes $\chi_4$. Scale bar: 20 $\mu m$.}
\end{figure}

To examine if the system is in an orientational glass state between $\phi_g^\theta$ and $\phi_g^T$, we measured the dynamical heterogeneity for rods with $p = 1.5$, at $\phi$ = 0.42, 0.81, and 0.83. The dynamical heterogeneity was quantified by the four-point dynamic susceptibility, $\chi_4(\Delta t)$ \cite{Berthier2005,Keys2007,Zhang2011} (see SI). $\chi_4$ is small at $\phi$ = 0.42 and 0.83 (Fig. 2(a)) signifying a relatively homogeneous dynamics, characteristic of a colloidal liquid and fully glassy phase, respectively. The homogeneous dynamics is also confirmed by fast-moving particles \cite{Weeks2000} (Fig. 2(b) and 2(d)). At $\phi = 0.81$, the $\chi_4$ for the translational motion displays a large peak, while for the orientational motion $\chi_4$ remains nearly flat without a distinct peak. This suggests that the heterogeneous dynamics is strong only in the translational motion, but not the orientational motion. This is the characteristic of an orientational glass state \cite{Zheng2011,Keys2007}.

\begin{figure}
\includegraphics[width=1.05\columnwidth]{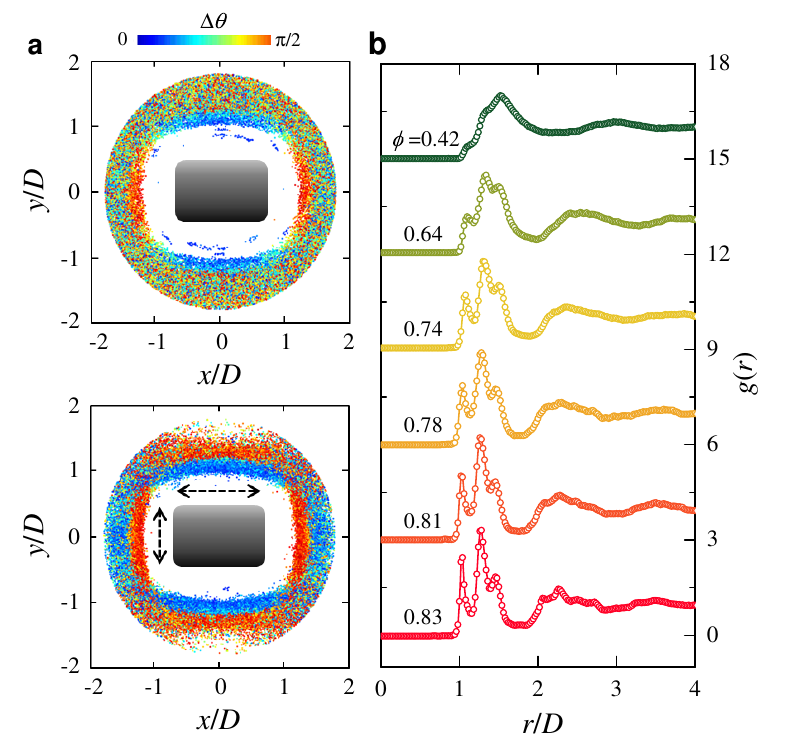}
\caption{\label{fig3} \textbf{Rods form local orientational structures approaching the rotational glass transition.} (a) The spatial distribution of the orientational configuration of neighboring rods ($\Delta\theta$) at $\phi$ = 0.42 (top panel) and 0.73 (bottom panel) for the rod monolayer with $p = 1.5$. Blue and red colors represent the parallel and perpendicular configurations, respectively. Arrows in bottom panel indicate that particles can keep its orientation while move along the axis of the neighboring particles. (b) The radial distribution function $g(r)$. For clarity, the curve has been shifted vertically for different $\phi$.}
\end{figure}

It was suggested that ellipsoids with the large aspect ratios ($p > 2.5$) can form pseudo-nematic domains, which in turn lead to the two-step glass transition \cite{Zheng2011}. In the rod systems here, for $p = 1.5$, no pseudo-nematic domains were observed for any $\phi$. For rods with larger anisotropy $p = 2.3, 2.7$, micro-domain structures with local nematic order can indeed be observed (see Fig.S5).  This observation suggests that the nematic domains are not the cause of the two-step glass transition, but rather a by-product of the underlying mechanism that drives the two-step glass transition.

To gain insights into the nature of the glass transition in this rod system, we look into the distribution of the relative angles between the rods.  This distribution must reflect the state of the colloidal matter just before it fell out of equilibrium and became stuck in a glassy state.  In Fig.3(a), the spatial distributions of the local orientational configuration for rods with $p = 1.5$ are shown in a graphical representation. The positions and orientations for all neighboring particles are normalized with respect to the reference rod sketched in the center of the figure, and the neighboring particles are plotted as the points. The color of the point denotes the relative orientation of the neighboring particle, $\Delta\theta$. A clear observation is that with increasing packing density, the rods are oriented relative to each other in either parallel or perpendicular local configurations (see blue and red color regions). Such increased local configurations are confirmed by the radial distribution function $g(r)$. With increasing $\phi$, the first peak in $g(r)$ splits into three peaks. These three peaks represent two parallel local configurations (the first peak and the third peak) and a perpendicular local configuration (the second peak) (Fig. 3(b)). The peaks become more pronounced with increasing $\phi$, indicating the enhancements of the parallel and perpendicular local configurations when approaching the glass transition.

We suggest that the distribution of the relative angles between nearest neighbors observed in the deep glassy phase is representative of the probability distribution between the neighboring rods prior to being immobilized. Since the detailed balance is likely to be preserved locally for the fluctuations between the rods just before they fall out of equilibrium during packing, thus one can use Boltzmann statistics to extract the effective free energy landscape of the rod-rod interactions from their relative angle distribution over the whole glass transition process.  It is worth noting that using local fluctuating configurations in a 2D colloidal lattice to extract effective potential was successfully attempted previously \cite{pertsin2001}.   After binning the data in Fig. 3a, we obtained the probability distribution of local orientational configuration, $P(\Delta\theta)$. The distribution shows a clear increase of the probability of the parallel and perpendicular configurations and an obvious decrease of the intermediate configurations as as $\phi$ increases (Fig.4(a)). Assuming $P(\Delta\theta) \propto e^{-V(\Delta\theta)/k_BT}$, where $V(\Delta\theta)$ is the effective potential energy (Gibbs free energy) of the local configuration $\Delta\theta$, $k_B$ is the Boltzmann constant and $T$ is the temperature. The estimated $V(\Delta\theta)$ from -$k_BTlnP$ is shown in Fig.4(b). Indeed the low energy configurations belong to the parallel or perpendicular local structure of the rods, and the angled configurations are less energy favorable.

\begin{figure}
\includegraphics[width=0.9\columnwidth]{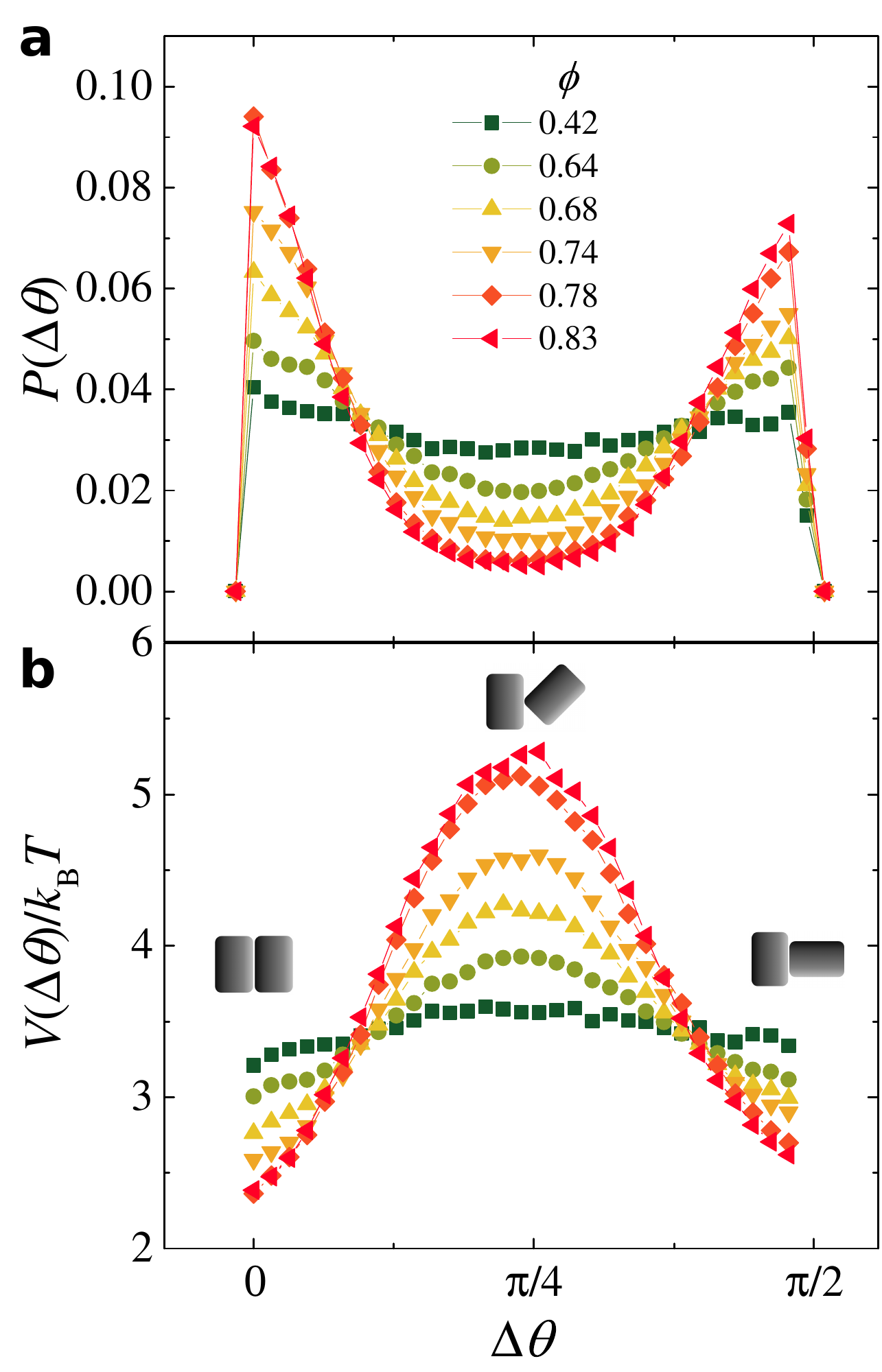}
\caption{\label{fig4} \textbf{Increase of concentration of the rods results in free energy barriers in orientational configurations.} (a) The probability of $\Delta\theta$ at different $\phi$ for rods with $p = 1.5$. (b) Potential energy $V(\Delta\theta)$ of the orientational configuration at different $\phi$. $V(\Delta\theta)$ was estimated from -$lnP(V(\Delta\theta))$. The relative arrangements of rods at three relative angles are shown.}
\end{figure}

Our local configuration analyses make it possible now to quantitatively address the structural origin associated with the two-step glass transition, by presenting the activation barrier as a function of packing density (Fig. 5). Interestingly, the barrier shows a plateau as the system enters into the orientational glass regime.  Further increasing density $\phi$ the barrier rises again when the system enters the second glass transition where the translational motion freezes. The plateau in the activation barrier is particularly striking. It suggests that the barrier to rotation in the orientational glass state is weakly dependent on the density.   Motivated by this puzzling effect, we repeated measurements on colloidal rods of two additional aspect ratios, $p$ = 2.3 and 2.7, the results are plotted together with $p$ = 1.5 in Fig.5. It is found that the plateau effect is present for all three values of aspect ratios.  Intuitively, one expects the barrier between the parallel and perpendicular configurations to be larger for particles of larger aspect ratios.  This seems to be true at low packing density, in the fully liquid phase, e.g. see Fig.5 at $\phi$=0.65.  The fact that the plateau of the activation barrier is smaller for particles of large aspect ratios can only be explained if the height of the plateau represents a threshold or critical barrier that forces the system out of equilibrium.  Since it takes longer time for longer rods to rotate relative to each other, thus the system will be stuck at a lower barrier height.  We thus propose a universal mechanism for the two-step glass transition of anisotropic particles: the large activation energy barriers between the local configurations hinder local rotation of anisotropic particles. This results in a decoupling between translational and orientational motion, consequently leading to a two-step glass transition.

\begin{figure}
\includegraphics[width=0.9\columnwidth]{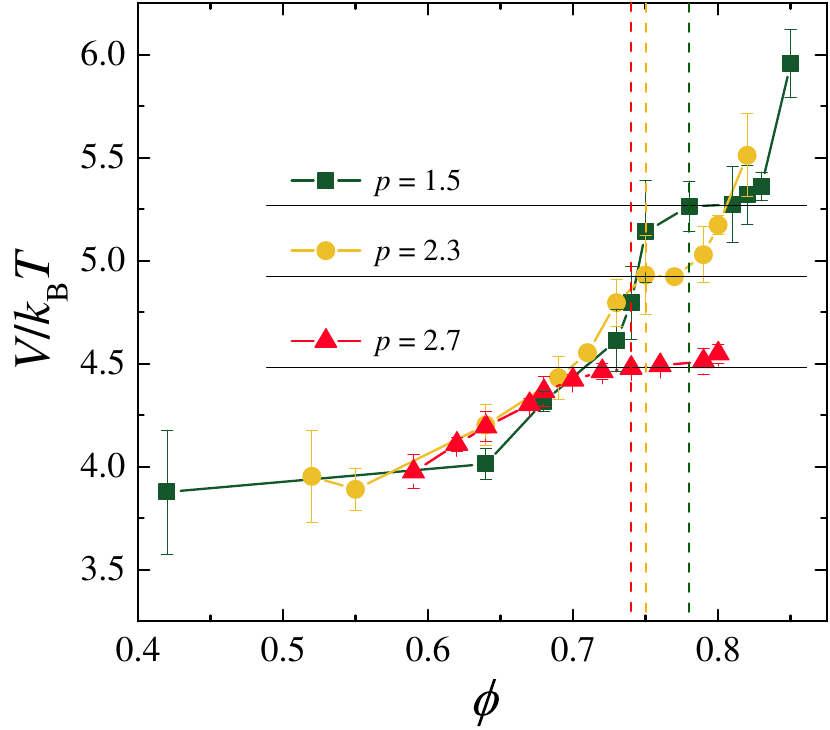}
\caption{\label{fig5} \textbf{Activation barrier increases sharply when approaching the orientational glass transition point.} The vertical lines indicate the orientational glass transition points for the rods with different aspect ratios, $p$. The transition points, are $0.74$ for $p = 2.7$, $0.75$ for $p = 2.3$, and  $0.78$ for $p = 1.5$. The horizontal lines indicate the heights of barrier plateaus.}
\end{figure}

It is interesting to compare our observations here with that in the two-stage melting of a 2D crystal \cite{Kosterlitz1973,Halperin1978,Young1979}. From the solid side, the first transition is edge dislocation unbinding, free edge dislocations allow parts of the lattice to slide past each other, making the effective shear modulus vanish at long length and time scales, rendering the system a liquid \cite{Kosterlitz1973}.  However, the orientational order remains, hence a hexatic phase \cite{Halperin1978}.  Only at a higher temperature when disclinations unbind, one part of the system can rotate freely relative to another, resulting in the isotropic liquid \cite{Halperin1978}. Here the situation is strikingly similar: starting from the frozen glass state, with lowering packing density, when the rods can slide past each other, it first gives rise to an orientationally frozen glass but a translationally flowing liquid \cite{Zheng2011}.  Only when density is further lowered, when the rods have free space to rotate, an isotropic liquid forms.  

We found that a two-step glass transition in a 2D colloidal suspension of rods during which there is no pseudo-nematic domain formation. Instead we observed that there exist locally favorable configurations due to shape anisotropy of the rods. Assuming that the distribution of the relative angles between the rods in the glassy state represents the local equilibrium configurations of the rods before they become stuck, one can invoke Boltzmann distribution to extract a local free energy distribution.  The existence of two preferred configurations implies a free energy barrier between them.  This barrier becomes more pronounced with increasing particle concentration as expected. We conclude that this energy barrier, which inhibits particle rotation, is the mechanism of a two-step glass transition in colloidal suspensions of rods. And we believe the energy barrier is the key to understand the two-step glass transition in other systems of anisotropic particles beyond rods and ellipsoids.

 \textbf{Acknowledgments:} We acknowledge helpful discussions with Yilong Han, J.M. Kosterlitz, and Paul Steinhardt. This work was financially supported by National Natural Science Foundation of China (Grant Nos. 11574222, 11704269, and 21522404), Natural Science Foundation of the Jiangsu Higher Education Institutions of China (Grant No. 17KJB140020), and the PAPD program of Jiangsu Higher Education Institutions.

\bibliographystyle{apsrev4-1}
\bibliography{Rebib}

\end{document}